\documentclass[oupdraft]{bio}
\usepackage[colorlinks=true, urlcolor=citecolor, linkcolor=citecolor, citecolor=citecolor]{hyperref}
\usepackage{amsthm}
\usepackage{amsmath}
\usepackage{algpseudocode}

\usepackage{natbib}
\setcitestyle{round}
\usepackage{natbibspacing}
\usepackage{pdflscape}
\usepackage{lscape}
\usepackage{enumitem}
\usepackage{color}
\usepackage{epsf}
\usepackage{algpseudocode}

\usepackage{graphics}
\usepackage{graphicx}
\usepackage{epsf}
\usepackage{epstopdf}
\usepackage{epsfig}

\def\beqnn{\begin{eqnarray*}}
\def\eeqnn{\end{eqnarray*}}
\def\beqlb{\begin{eqnarray}}
\def\eeqlb{\end{eqnarray}}

\begin{document}

% Title of paper
\title{Multiple Comparison Procedures for Neuroimaging Genomewide Association Studies}
% List of authors, with corresponding author marked by asterisk
\author{Wen-Yu Hua$^{\dagger}$, Thomas E. Nichols$^{\ddagger}$, Debashis Ghosh$^{\dagger}$, for the Alzheimer's Disease Neuroimaging Initiative$^{\ast}$\\[4pt]
% Author addresses
\textit{$^{\dagger}$ Department of Statistics, Penn State University,\\[-5pt]
$^{\ddagger}$ Department of Statistics, University of Warwick }\\[-5pt]
% E-mail address for correspondence
{wxh182/ghoshd@psu.edu}}

% Running headers of paper:
\markboth%
% First field is the short list of authors
{Hua, Nichols, and Ghosh}
% Second field is the short title of the paper
{Multiple Comparison Procedures for Neuroimaging Genomewide Association Studies}

\maketitle

% Add a footnote for the corresponding author if one has been
% identified in the author list

\begin{abstract}
{Recent research in neuroimaging has focused on assessing
associations between genetic variants that are measured on a genomewide scale
and brain imaging phenotypes. A large number of works in the area apply massively
univariate analyses on a genomewide basis to find single nucleotide polymorphisms
that influence brain structure. In this paper, we propose using various
dimensionality reduction methods on both brain structural MRI scans and genomic
data, motivated by the Alzheimer's Disease Neuroimaging Initiative (ADNI) study.
We also consider a new multiple testing adjustment method and compare it with two existing false discovery rate (FDR) adjustment methods.
The simulation results suggest an increase in power for the proposed method. The real data analysis suggests that the proposed procedure is able to find associations between genetic variants and brain volume differences that offer potentially new biological insights.}
{Distance covariance ; Genomewide association studies; Local false discovery rate; Multivariate analysis; Neuroimaging analysis; Positive false discovery rate.}
\end{abstract}

\footnotetext{Data used in preparation of this article were obtained from the Alzheimer's Disease Neuroimaging Initiative
(ADNI) database (adni.loni.ucla.edu). %As such, the investigators within the ADNI contributed to the design
%and implementation of ADNI and/or provided data but did not participate in analysis or writing of this report.
%A complete listing of ADNI investigators can be found at \href{http://adni.loni.ucla.edu/wp-content/uploads/how_to_apply/ADNI_Acknowledgement_List.pdf}{ADNI Acknowledgement List}.
}

\maketitle

\section{Introduction}
\label{sec1}

Advanced automated image processing techniques have allowed the assessment of the genetic association with brain phenotypes for complex diseases, such as schizophrenia \citep{Potkin09}, and Alzheimer's disease \citep{Furney2010}. In this work, we consider data from the Alzheimer's Disease Neuroimaging Initiative (ADNI) project \citep{ADNI03} consisting of genetic variants encoded as single nucleotide polymorphisms (SNPs) across whole genome, and brain volume size measured by tensor-based morphometry (TBM) based on structural magnetic resonance imaging (MRI) scans. Specifically, TBM computes the volume of a local brain region in a given subjects' MRI relative to an average template image based on healthy subjects. Since a signature of Alzheimer's Disease (AD) is the thinning of cortical gray matter and an increase of cerebral spinal fluid volume (particularly in the ventricles), TBM is sensitive to AD-related changes through decreases in volume of the cortex, and increases in volume of the ventricles. Therefore, the goal of this work is to find the genetic variants that result in change of brain volumes.

\citet{Stein10} conducted a voxelwide and genomewide study using TBM maps from each subject, where each voxel is evaluated with a regression at each SNP based on the SNP's minor allele count, and using demographic variables as features with quantitative trait as responses. In their experiment, no significant loci were found after a false discovery rate based on the multiple testing adjustment procedure at level $0.05$. In a later study, \citet{Stein10b} performed a genomewide search on two brain phenotypes (temporal lobe and hippocampal volume) based on the prior results from the literature. To investigate the associations, they collected an independent sample for each phenotype and performed adjusted regression analysis on the baseline population. Overall, two significantly associated SNPs were identified: $rs10845840$, located on chromosome 12 within an intron of the $GRIN2B$ gene, and $rs2456930$, which is in an intergenic region of chromosome 15. Both SNPs were significantly associated with bilateral temporal lobe volume, while no significant SNPs were found to have associations with hippocampal phenotype.

For any univariate approach to analysis, multiple testing procedures should be employed as there are many statistical tests being considered simultaneously. A recent error quantity called false discovery rate (FDR) was proposed for the multiple comparisons problem by \citet{Benjamini95}. Later, \citet{Storey02} and \citet{Storey03} defined the positive false discovery rate ($p$FDR) that is the conditional expectation of false positive findings given at least one positive identifications has occurred, and also proposed a $q$-value algorithm to control the $p$FDR. \citet{Efron01} defined a local false discovery rate (locfdr), a Bayesian version of FDR. For its estimation, they fit a mixture model to a Gaussian transformation of the inverse cumulative distribution of the $p$-values. To relate the frequentist and Bayesian versions of FDR, \citet{Efron01}, \citet{Efron02}, and \citet{Storey02} proved that the FDR controlled by the Benjamini and Hochberg procedure is equivalent to empirical Bayesian FDR given the rejection regions. Furthermore, \citet{Michael03} proposed a hierarchical mixture of Gammas for the multiple comparisons problem and \citet{Omkar10} showed that the locfdr estimation controls FDR/$p$FDR over the entire exponential distribution family.

The previously published ADNI analyses were able to find associated SNPs or genes that are likely to be related to some specific voxels of the brain scans. However, neighboring structures of the brain were not being considered, and this information could play an important role in associations with disease risk. In this work, this issue is addressed by combining the neighboring voxels into 119 regions based on the GSK CIC atlas \citep{Tziortzi11}, and then the effects on the regions are simultaneously assessed using the distance covariance statistic \citep{Szekely07}, which allows for inference on the relationship between a 119-dimensional multivariate phenotype and a single SNP predictor across the entire genome.

We make two contributions to the analysis of the ADNI neuroimaging genomewide study. First, we utilize distance covariance for the analysis of genomewide association study. This framework is able to establish the relationships between genomic variants and brain structural MRI where the entire brain is a multivariate response. By considering a multivariate response variable, we reduce the number of tests being done relative to an approach such as in \citet{Stein10}, which results in more powerful inference. Second, we propose a local fdr modeling algorithm to address the multiplicity which is to fit a two-component mixture of Gammas on the distance covariance statistics. One probabilistic output of this model is the local fdr. This leads to a decision-theoretic rule for selecting significant SNPs that is related to the approach of \citet{Michael03}. In the multiple testing step, we also evaluate two existing methods for comparison. Based on our simulation studies and real data analysis using ADNI, experiments show that the proposed method is able to control FDR at different $\alpha$ levels as well as provide more powerful findings than \citet{Stein10}'s work. In addition, we also present the pathway analysis based on our significant findings in the supplementary material, and show that the significant SNPs survived from our procedures provide signal enrichment functions through pathway to AD from the Database for Annotation, Visualization and Integrated Discovery (DAVID).
%
%The structure of this paper is as follows: Section \ref{sec2} reviews the details of ADNI data. The distance covariance statistic and the multiple testing adjustment procedure are introduced in Section \ref{sec3}. In Section \ref{sec4}, we present the results and discuss the performances with comparing to two existing methods. Finally, some discussion concludes Section \ref{sec5}, and a short description of the supplementary is in Section \ref{sec6}, with Section \ref{sec7} containing the acknowledgements.

\section{Materials}
\label{sec2}
Data used in the preparation of this article were obtained from the ADNI study \citep{ADNI03}. The SNP data and the TBM data from the ADNI study are processed by Paul Thompson's group, which are the same as those used in the previous studies \citep{Stein10}. For the sake of completeness, we describe the genetic and imaging data preprocessing in the following section.

\subsection{ADNI study}
The ADNI was launched in 2003 by the National Institute on
Aging (NIA), the National Institute of Biomedical Imaging and Bioengineering (NIBIB), the Food and Drug
Administration (FDA), private pharmaceutical companies and non-profit organizations, as a 60 million, 5-
year public-private partnership. The primary goal of ADNI has been to test whether serial MRI, positron emission tomography (PET), other biological markers, and clinical and neuropsychological assessment can be combined to measure the progression of mild cognitive impairment
(MCI) and early AD. Determination of sensitive and specific markers of very early AD
progression is intended to aid researchers and clinicians to develop new treatments and monitor their
effectiveness, as well as lessen the time and cost of clinical trials. The Principal Investigator of this initiative is Michael W. Weiner, MD, VA Medical Center and University of California San Francisco. ADNI is the result of efforts of many co-investigators from a broad range of
academic institutions and private corporations, and subjects have been recruited from over 50 sites across
the U.S. and Canada. The initial goal of ADNI was to recruit 800 subjects but ADNI has been followed by
ADNI-GO and ADNI-2. To date these three protocols have recruited over 1500 adults, ages 55 to 90, to
participate in the research, consisting of cognitively normal older individuals, people with early or late MCI,
and people with early AD. The follow up duration of each group is specified in the protocols for ADNI-1,
ADNI-2 and ADNI-GO. Subjects originally recruited for ADNI-1 and ADNI-GO had the option to be followed
in ADNI-2. For up-to-date information, see www.adni-info.org.

Of the 852 total subjects released by the ADNI dataset, the availabilities of both brain structural MRI and genetics records were found in 741 subjects. The data for these subjects are used for our experiments, where the volumetric brain differences are assessed in 206 normal older controls, 358 MCI subjects, and 177 AD patients.

\subsection{Genetic analysis}
ADNI released 620,901 SNPs using the Illumina 610 Quad array. SNPs that did not fulfil the following quality control criteria were excluded: genotype call rate smaller than $95\%$, significant deviation from Hardy-Weinberg equilibrium where $p$-value $ <5.7 \times 10^{-7}$, allele frequency smaller than 0.10, and a quality control score of smaller than 0.15. After applying this list of quality criteria, we obtain a total of 448,244 SNPs for the analysis. The number of SNPs measured on each chromosome is in table 1 in our supplementary material.

\subsection{Brain MRI scans}
\label{mri}
Three-dimensional $T_1$-weighted baseline MRI scans were analyzed using TBM: a method for representing structural differences between local brain regions and a template brain into a deformation field \citep{Friston04}. The deformation field contains the information on relative positions of different brain scans, while the local shapes (such as volumes, lengths and areas) are encoded in the Jacobian matrix. Therefore, TBM can be used to recognize the local shape of brain differences. The MRI scans were acquired at 58 different ADNI sites, all with 1.5T MRI scanners using a sagittal 3D MP-RAGE sequence for across-site consistency \citep{Jack08}. All images were calibrated with phantom-based geometric corrections. The scans were linearly registered with 9 parameters to the International Consortium for Brain Image template \citep{Mazziotta01} to adjust for differences in brain position and scaling. Each subject's MRI scan was registered against a template scan which is the average of all the healthy subjects (minimal deformation template), using a non-linear inverse-consistent elastic intensity-based registration method \citep{Leow05}. Furthermore, voxel size variation from registration is represented as the voxel intensity, which is the volumetric difference between the subject and the reference template, calculated from taking the determinant of the Jacobian matrix of the deformation fields. Finally, each brain scan volume is down-sampled to $1/4$ of its original size (using trilinear interpolation to $4 \times 4 \times 4$ $mm^{3}$), which results into $31,622$ total voxels per scan for faster experimental processing. Similar to \citet{Stein10}, we use the volumetric difference representation of MRI as the quantitative measure of brain tissue volume difference for the genomewide association analysis.

\begin{figure}[!p]
\centering
\includegraphics[width=0.8\linewidth]{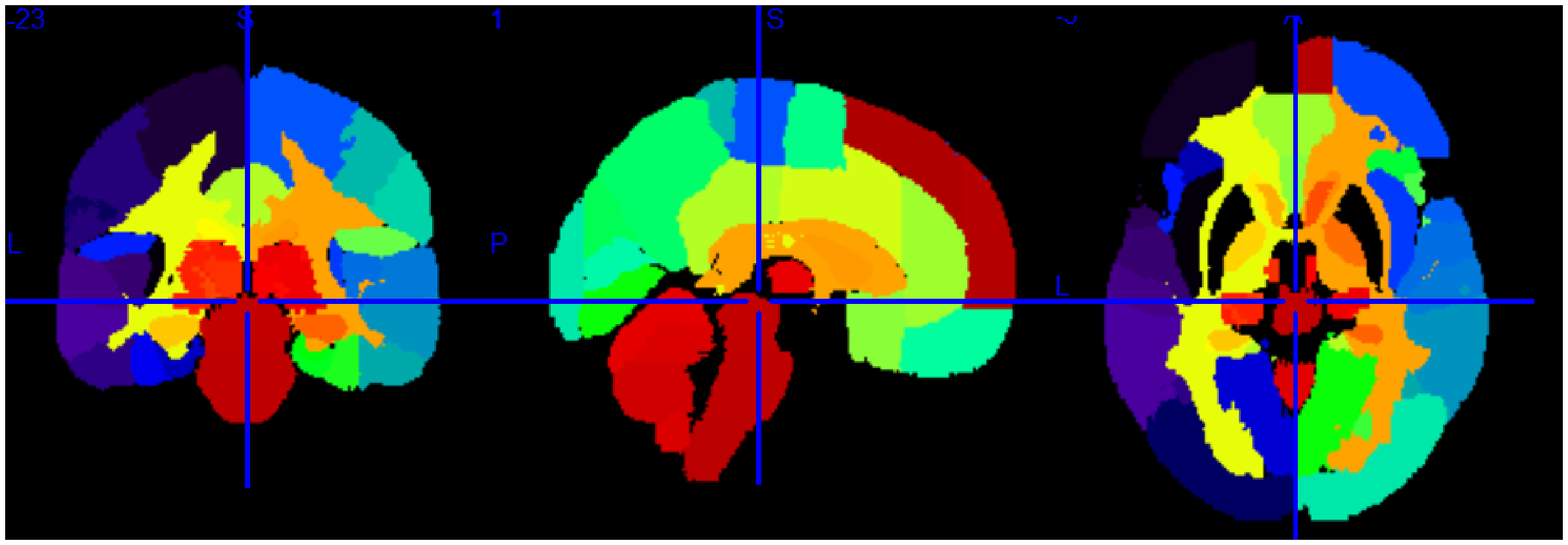}
\caption{\emph{Plots showing from left to right: Coronal, Sagittal and Axial views of GSK CIC Atlas, color coded by the 119 region of interests.}}
\label{fig:fsl}
\end{figure}

We explore genomewide associations with brain volume difference in terms of voxels; we also perform the same analysis based on groups of voxels, which is the focus of this work. This region of interests (ROIs) approach is a type of dimensionality reduction method that allows for information on local neighborhoods of voxels to be pooled, and reduces possible noise that associates with performing analysis using the entire brain voxels, we denote this as the region-wide study. In order to conduct the experiment using 119 ROIs, we extracted voxels from each brain region, and computed the average Jacobian scores (per region) that make up the 119 different brain regions from the GSK CIC Atlas as shown in Fig. \ref{fig:fsl}, which is based on the Harvard-Oxford atlas with a 6-level hierarchy. To extract the corresponding voxels from each brain region in the atlas, we used the FLIRT linear registration tool from FSL (\citet{Jenkinson01}, \citet{Jenkinson02}, \citet{Smith04}, and \citet{Woolrich09}) in order to register the brain atlas to our template scan. This allows us to extract voxels of different brain regions from the subject's scan and the registered atlas by direct comparison. We then used the average per-region Jacobian scores from each of the 119 ROIs as the response into genomewide association.

\section{Methods}
\label{sec3}
\subsection{Distance covariance}
\label{Association analysis}
The work of distance covariance in \citet{Szekely07} and \citet{Szekely09} is discussed here. Let $\phi_X$ and $\phi_Y$ be the characteristic functions of $X$ and $Y$, where $X \in \mathbb{R}^p$ and $Y \in \mathbb{R}^q$ are two random vectors from two arbitrary dimensions $p$ and $q$, respectively. The distance covariance $\mathrm{dCov}^2(X,Y)$ between random vectors $X$ and $Y$ is a non-negative value with finite first moments:
\beqlb
\label{dCov}
\mathrm{dCov}^2(X,Y)&=&\|\phi_{X,Y}(x,y)-\phi_X(x)\phi_Y(y)\|^2  \\
&=& \int_{\mathbb{R}^{p+q}} \! |\phi_{X,Y}(x,y)-\phi_X(x)\phi_Y(y) |^2 w(x,y) \, \mathrm{d} x \mathrm{d} y,\nonumber
\eeqlb
where $w(x,y)$ is a positive weight function for which the integral in Eq. (\ref{dCov}) exists.

The sample distance covariance estimator from \citet{Szekely07} and \citet{Szekely09} requires that there be no missing values among observations $X_i$'s and $Y_j$'s for $i,j=1,...,n$. In order to relax this requirement, we propose a modified version by assuming the data is missing completely at random (MCAR, \citet{heitjan1996distinguishing}).
 %The main idea of this modification is to save additional computational cost when modeling the missing values.
Here, $\delta$ is defined as an indicator which indicates if a variable is missing or present:
\beqlb
\label{delta}
\delta_k&=& \begin{cases} 1, & \mbox{if variable $k$ is present}\\ 0, & \mbox{if variable $k$ is missing} \end{cases}.
\eeqlb
Adjusting the indicator $\delta$ for observations $X_i$'s and $Y_j$'s puts larger weights on observations with no missing values and zero weight on observations with missing values. For $i,j=1,...,n$, Our modified preliminary statistics according to \citet{Szekely07} as $A'_{ij}= a'_{ij}-\bar{a'}_{i.}-\bar{a'}_{.j}+\bar{a'}_{..}$, where
\beqlb
\label{bigAprime}
a'_{ij} = \frac{|X_i-X_j|_p\delta_i\delta_j}{P(\delta_i=1)P(\delta_j=1)}
\eeqlb
 and
\beqnn
\bar{a'}_{i.} = \frac{1}{n}\sum^n_j a'_{ij} \ \ \ \
\bar{a'}_{.j} = \frac{1}{n}\sum^n_i a'_{ij} \ \ \ \
\bar{a'}_{..} = \frac{1}{n^2}\sum^n_{i,j} a'_{ij}.
\eeqnn
Similarly, we define $B'_{ij}= b'_{ij}-\bar{b'}_{i.}-\bar{b'}_{.j}+\bar{b'}_{..}$ with its elements taking the same form as $A'_{ij}$.
%$$b'_{ij}=\frac{|Y_i-Y_j|_q\delta_i\delta_j}{P(\delta_i=1)P(\delta_j=1)}$$

The modified sample distance covariance $\widetilde{\mathrm{dCov}}^2_n(X,Y)$ is then given by
$\widetilde{\mathrm{dCov}}_n^2(X,Y) =n^{-2} \sum^n_{i,j} A'_{ij}B'_{ij}$.
Having proposed a modified empirical distance covariance for situations where missing values are present, we can study its asymptotic property under the independent assumption. The expectation of $a'_{ij}$ in Eq. (\ref{bigAprime}) is:
\beqlb
\label{expectation}
E(a'_{ij})&=&E \left \{ \frac{|X_i-X_j|_p\delta_i\delta_j}{p(\delta_i=1)p(\delta_j=1)} \right \} \\
&=&E(|X_i-X_j|_p) \nonumber
%&=& E(a_{ij}) \nonumber
\eeqlb
and similarly, $E(b'_{ij})=E(|Y_i-Y_j|_q)$.
Arguing as in \citet{Szekely07}, we have that if $E|X|_p < \infty$ and $E|Y|_q < \infty$, then $\widetilde{\mathrm{dCov}}_n \rightarrow_{a.s} \mathrm{dCov}$. Consequently, it can be shown that
\beqlb
\label{T}
T = n \times \widetilde{\mathrm{dCov}}^2_n/ T'_2 \rightarrow_{D} Q,
\eeqlb
where $T'_2=n^{-2}\sum^n_{i,j} a'_{ij}n^{-2} \Sigma^n_{i,j} b'_{ij}$ and $Q$ is a positive semidefinite quadratic form of centered Gaussian random variables with $E(Q)=1$. %This property can help us to identify both non-linear or non-monotone dependencies between $X$ and $Y$.
\citet{Szekely07} proposed a permutation test for hypothesis testing. However, the permutation scheme is extremely computationally expensive when dealing with large scale data such as our genomewide association study. In terms of obtaining $p$-values, we apply a Gamma approximation for inference on the distance covariance statistics \citep{Gretton08}, which is discussed in Section \ref{Multiple testing procedure}.

For comparison purposes, we also investigate the case of missing at random (MAR) for imputing the missing values using a publicly available software PLINK \citep{Purcell07}. The results from the genotype imputation are addressed in Section \ref{sec5}.

\subsection{Multiple testing procedure}
\label{Multiple testing procedure}

We now review the multiple testing problem and define FDR. Assume that there are $m$ tests for the study, the goal is to identify the significant SNPs at a certain $\alpha$ level. Table \ref{tab:mtests} shows the possible outcomes of conducting $m$ tests simultaneously, for which the null hypothesis is true in $m_0$ of them. Of the $m$ tests of hypotheses, $W$ hypotheses are failed to be rejected, and $R$ rejected the null hypothesis.

\citet{Benjamini95} introduced a new measure called FDR, defined as:
\beqlb
FDR&=& E \left [\frac{V}{R}|R>0 \right ] P(R>0)
\eeqlb
\citet{Storey02} and \citet{Storey03} proposed another measure, positive false discovery rate ($p$FDR), which is the expected false-positive rate conditioned on positive finding ($P(R=0)>0$). The $p$FDR takes the following form:
\beqlb
p\mbox{FDR} &=& E\left [\frac{V}{R}|R>0\right ]
\eeqlb
Our aim is to control $p\mbox{FDR}$, and we present three algorithms to achieve this goal for the remainder of this section.

The first algorithm is the $q$-value algorithm, which was first presented by \citet{Storey02}. $q$-value requires that the prior knowledge of the null distribution of the test is known, such that the $p$-values can be computed under the null density. In the case of distance covariance, \citet{Gretton08} proposed to fit a Gamma distribution as the null density, with the following parameters:
\beqlb
\label{moment}
\alpha=\frac{E^2(T)}{Var(T)}   \ \ \ \ \beta = \frac{Var(T)}{E(T)},
\eeqlb
where $T$ is defined as Eq. (\ref{T}). Hence, the parameters in Eq. (\ref{moment}) can be estimated by the distance covariance statistics, and the $p$-values are able to be computed from the Gamma approximation for the $q$-value method. The algorithm for the $q$-value method is as follows: first, for each $T_i$, we compute the $p$-value $p_i$ under the Gamma approximation; we then compute $q$-values $q_1,...,q_m$ for each test using the method of \cite{Storey03}; by defining $\tilde{q}=\arg\max_{i} \{q_i \leq \alpha\}$, we reject all tests with $q_i \leq \tilde q$.
%\end{enumerate}
%
%\noindent \textbf{Algorithm 1}:\\
%Input: $m$ hypotheses with statistics $T_1,...,T_m$ in Eq. (\ref{T}).
%\begin{enumerate}[noitemsep,nolistsep]
%\item For each $T_i$, compute the $p$-value $p_i$ under the Gamma approximation.
%\item Compute $q$-values $q_1,...,q_m$ for each test using the method of \cite{Storey03}
%\item Define $\tilde{q}=\arg\max_{i} \{q_i \leq \alpha\}$, and reject all tests with $q_i \geq \tilde{t}$.
%\end{enumerate}
\citet{Storey02} and \citet{Storey03} have showed that the $q$-value algorithm (Algorithm 1) controls FDR under the desired $\alpha$ level. Note that step 2 of the above algorithm is computed using the publicly available R-package \textsf{qvalue}.

The second algorithm uses local fdr (Eq. (5.1) in \citet{Efron01}) to control $p$FDR. In \citet{Efron01}'s work, the null distribution is assumed either known or collected by permutation. Here, we chose to use a Gamma approximation with empirical estimations of Eq. (\ref{moment}) as the null density candidate for distance covariance statistics, and the detailed derivations are presented in Section 2 of the supplementary materials. The following is a summary of \citet{Efron01}'s work.\\
\noindent \textbf{Algorithm 2}:\\
Input: $m$ hypotheses with statistics $T_1,...,T_m$ in Eq. (\ref{T}).
\begin{enumerate}[noitemsep,nolistsep]
\item For each $T_i$, compute the $p$-value $p_i$ under the Gamma approximation.
\item For each $p_i$, $z_i=\Phi^{-1}(p_i)$, for all $i$.
\item Estimate the parameters using a Gaussian mixture model (GMM) to the $z_i$'s.
\item Compute local fdr as defined by Eq. (5.1) in \citet{Efron01}.
\item $\widehat{p\mbox{FDR}}$($z$) is the conditional expectation of local fdr given $z \in \Gamma$.
\item Define $\tilde{z}=\arg\max_{z} \{\widehat{p\mbox{FDR}}(z) \leq \alpha\}$, and reject all tests with $z_i$ in rejection region. In theory, this gives an $p$FDR no greater than $\alpha$.
\end{enumerate}

We now propose a new algorithm (denoted as local fdr modeling) for multiple testing adjustment. The traditional multiple correction methods are based on $p$-values (e.g., algorithm 1 and 2), while our proposed method models the test statistics directly. The algorithm for the local fdr modeling is similar to algorithm 2, but skipping the second step. This rule is similar to the one proposed by \citet{Michael03} in a different genomics setting, where more powerful inference can be obtained by not mapping the test statistics from $t_i$'s to $z_i$'s.\\% The details of our proposed algorithm 3 can be found in the supplementary materials (local fdr modeling algorithm).\\
\noindent \textbf{Algorithm 3: (local fdr modeling)}:\\
Input: $m$ hypotheses with statistics $T_1,...,T_m$ in Eq. (\ref{T}).
\begin{enumerate}[noitemsep,nolistsep]
\item Fit a two-component mixture of Gammas to $T_1,...,T_m$.
\item Compute local fdr as defined by Eq. (5.1) in \citet{Efron01}.
%\item $\widehat{p\mbox{FDR}}$(t) is the conditional expectation of $\widehat{fdr}(t)$ given $t \in \Gamma$ based on theorem \ref{average} (the average theorem in \citep{Efron01}).
\item Define $\tilde{t}=\arg\max_{t} \{\widehat{p\mbox{FDR}}(t) \leq \alpha\}$, and reject all tests with $t_i \geq \tilde{t}$. Where $\widehat{p\mbox{FDR}}$(t) is the conditional expectation of local fdr given the rejection region.
\end{enumerate}
\section{Simulation study and real data analysis}
\label{sec4}
We have implemented distance covariance in Matlab for our experiments. The R packages \textsf{qvalue}, \textsf{mixfdr}, and \textsf{mixtools} were used for the multiple testing procedures. All the analyses were accomplished by using the university high performance computing cluster, which consists of 128 Intel Xeon E5450 nodes, each with 8 cores and 32 GB of memory.

\subsection{Simulation Design}
\label{Simulation Design}

To evaluate the methods described in Section \ref{sec3}, we simulated the data to examine the FDRs and power estimates by controlling $\alpha$ at desired levels, and the settings of the simulation study were to mimic the structure of the genotypes and the phenotypes of the ADNI study. We considered two types of correlations (i.e., the pure linear correlation, and the mixed linear and non-linear correlations) and the impact of univariate and multivariate effects into three simulation settings. For each setting, the samples were generated from a null and an alternative population, and 1000 genotypes were generated for multiple testing. Then, we examined the association one genotype at a time across the 1000 genotypes for the following three settings. In this first case, we generated 50 paired samples: each pair included a single genotype and a phenotype, and followed a bivariate Normal distribution, where the correlation coefficient $\rho$ was $0.8$ under the alternative or around zero under the null. For the second and the third case, the sample size was 100, and the univariate genotype was generated from $N(0,1)$ while the dimensions of phenotype were enlarged to 30. The $100 \times 30$ phenotype data formed the mixed association effects between phenotypes and the genotypes under the alternative, where the mixed associations were linear, exponential and quadratic transformations (i.e., 10 duplicated copies of 100 genotypes; 10 exponential transformations of 100 genotypes; and 10 quadratic forms of 100 genotypes). For the null population, the single genotype again was generated from $N(0,1)$ and the 30-dimensional phenotypes followed a multivariate Normal with mean 0 and covariance matrix $\Sigma$, where $\Sigma$ was independent in the second simulation design and positive dependent (diagonal terms are one and off diagonal terms are 0.5) on third case. The ratio between the null and alternative population was 19:1, and a total of 1000 runs were repeated for each setting to assess the FDRs and power performances.

\subsection{Simulation Results}
\label{simulation results}
Three FDR procedures were presented in Section \ref{Multiple testing procedure}, which we summarize again in the following:
\begin{description}[noitemsep,nolistsep]
\item[Alg. 1:] $p$-values (from Gamma approximation) + $q$-value method \citep{Storey03}
\item[Alg. 2:] $p$-values (from Gamma approximation) + local fdr method \citep{Efron02}
\item[Alg. 3:] local fdr modeling proposed in Section \ref{Multiple testing procedure}
\end{description}

Before the discussion of the FDRs and power estimates of three algorithms, we performed size analysis to evaluate if a Gamma approximation \citep{Gretton08} is a proper null density for algorithm 1 and 2. We generated 1000 (genotypes v.s. phenotypes) samples for the size analysis, where the associations were all from the null population for the three simulation settings; 50 runs were repeated to calculate the size. Table \ref{tab:sizeanalysis} reports the size estimates according to nominal values from 0.1, 0.2,...,1, and the size estimates are very close to their corresponding nominal values for all three simulations. Therefore, we concluded that the Gamma approximation is an appropriate null distribution for the distance covariance statistic.

Table \ref{tab:pFDRRresult} shows the average FDRs, the average powers, and their standard errors at nominal $\alpha$ levels 0.05, 0.1, 0.15 and 0.2 for the three simulations. The results show that the average FDRs are all close or lower to the desired $\alpha$ values. The powers of algorithm 2 and 3 outperform algorithm 1 for all $\alpha$ values; this implies the algorithms which utilize the local fdr method result in powerful inference. In addition, the average estimated power of simulation 3 is smaller but close to the power of simulation 2 at each level. This shows that the results of all multiple testing adjusted algorithms are slightly affected by the noise of the dependent covariance structure, but the overall performances are robust. Furthermore, the results of algorithm 2 and 3 are similar in our simulation studies, and this suggests that algorithm 3 controls FDR well.

\subsection{Application to ADNI data sets}

We evaluated the three algorithms using the ADNI dataset. For each test, the independent variable is a single SNP across the whole genome (448,244 SNPs). The multivariate response is a 119 dimensional vector (i.e. 119 ROIs), with each value corresponding to the average voxel value for such brain region, based on the GSK CIC Atlas. We also considered the entire brain imaging voxels (31,622 voxels) as another multivariate response for ADNI study, and the results are shown in supplementary materials. In addition to the three algorithms described in Section \ref{Multiple testing procedure}, we also implemented a modified version of \citet{Stein10}'s work, in which they originally considered simple linear regression (slr) as the association test between a single SNP and brain a voxel, with our modification being a single SNP and a brain region. For this method, we selected the brain region with the highest $p$-value at each SNP, then use the local fdr method to perform multiple testing adjustment. This procedure is denoted as algorithm 4 (slr+local fdr method).

Table \ref{tab:ADNIresult} displays the number of significant SNPs controlled by the $\alpha$ values from each algorithm. Notice that there were 1180 significant SNPs with $\alpha$ at 0.5 in algorithm 4 \citep{Stein10}, while algorithm 2 and 3 resulted in more than 20,000 findings, with algorithm 1 yielded slightly above 5,000 SNPs at $\alpha$ level 0.05. In order to compare the inference information of the significant SNPs in Table \ref{tab:ADNIresult}, the top 1180 SNPs were selected from each of the algorithm as the input variables for disease status classification. Specifically, we performed binary disease status classification (206 normal patients against 177 AD's) due to the fact that AD is the definitive form of the illness with much higher severity than MCI. We used LIBSVM \citep{CC01a} for binary classification with leave-one-out to compute the prediction accuracy. The majority count was 53.786\%, and the prediction accuracy of top 1180 SNPs from algorithm 1, 2 and 3 were all 57.441\%, as the top 1180 SNPs from the three algorithms were exactly the same. The prediction accuracy of algorithm 4 was the same as the majority count. In addition, algorithms 1, 2 and 3 at $\alpha$ level 0.05 found 5,388, 27,965, and 23,128 significant SNPs (table \ref{tab:ADNIresult}), and these SNPs yielded 57.964\%, 62.141\%, and 62.402\% prediction accuracies, respectively.
We have also performed the functional annotation clustering analysis using DAVID v6.7 \citep{DAVID03}. Table \ref{tab:David} lists the top 8 clusters enrichment scores and the total enrichment scores. Since the top 1180 SNPs from algorithms 1, 2 and 3 were identical, the enrichment scores from these three algorithms were also the same, with each having a total score of 10.257 which is greater than the total enrichment score of 5.739 from algorithm 4.

The above analyses imply that algorithm 4 \citep{Stein10} yields less significant findings even with a higher nominal $\alpha$ level, and the 1180 SNPs contain less information in both disease status classification and functional annotation clustering analysis. We further investigate the functional enrichment terms of algorithm 2 and 3 at $\alpha$ level 0.05 in the region-wide study, and the results are listed in supplementary materials.

\section{Discussion and Conclusion}
\label{sec5}

In this work, we have performed neuroimaging genomewide association studies using the ADNI dataset. The proposed method using distance covariance is able to identify the dependencies between the SNP variants and the brain volume differences, and utilize brain region interaction effects at the same time. We also proposed a local fdr modeling strategy and compared the performances with two existing multiple testing adjustment methods. The simulation studies showed that $p$-values computed from Gamma approximation with the local fdr method (algorithm 2) and local fdr modeling (algorithm 3) were able to control FDR at the proper levels. In the real data application, the significant SNPs found by distance covariance contained more information than simple linear regression \citep{Stein10} in both disease status classification, and function annotation clustering analysis. This is because simple linear regression only captures linear relationship between SNPs and brain MRI scans, while distance covariance is able to model non-linear associations.

In addition to the distance covariance statistic in Eq. (\ref{bigAprime}) that we have proposed for missing data, another option to deal with the missing values is to impute the genotypes by assuming the missing values are MAR. We used PLINK to impute the missing values in the ANDI study under the assumption of MAR, as it is computationally efficient \citep{Yun09}. The PLINK algorithm uses the standard EM algorithm and performs probabilistic estimation for each allele combination based on the relatively small regions of genome for each individual \citep{Purcell07}. Based on the results of our PLINK imputation, the non-missing rate of the data increased from 99.61\% to 99.67\%. Therefore, we work with the original datasets and assume MCAR. Exploring combinations of imputation algorithms with distance covariance measures deserves further investigation, but is beyond the scope of the paper.

There remain many open questions that could lead to important further developments. We utilized distance covariance to measure the relationship between genetic variants and differences in brain volumes in the first stage. This representation can be applied to capture the non-linear dependencies between two sets of vectors with arbitrary dimensions, but it might also suffer a possible bias when the number of dimensionality is much greater than the sample size \citep{cope10}. Therefore, we placed more emphasis of our results on the region-wide study in this work, and we plan to study regularization approaches to the dependency measure to reduce this bias in future work. It would also be desirable to develop distance covariance-type measures that explicitly incorporate the discrete nature of the SNP data.

\section{Supplementary Materials}
\label{sec6}
The structure of supplementary materials is as follows: Section 1 lists the number of SNPs for each chromosome, and Section 2 presents the details of algorithm 1, 2 and 3. The additional analyses of the ADNI data are shown in Section 3.

\section{Acknowledgments}
\label{sec7}
This research is supported in part by National Science Foundation grant DBI-1262538.
Data collection and sharing for this project was funded by the Alzheimer's Disease Neuroimaging Initiative
(ADNI) (National Institutes of Health Grant U01 AG024904). ADNI is funded by the National Institute on
Aging, the National Institute of Biomedical Imaging and Bioengineering, and through generous contributions
from the following: Abbott; Alzheimer's Association; Alzheimer's Drug Discovery Foundation; Amorfix Life
Sciences Ltd.; AstraZeneca; Bayer HealthCare; BioClinica, Inc.; Biogen Idec Inc.; Bristol-Myers Squibb
Company; Eisai Inc.; Elan Pharmaceuticals Inc.; Eli Lilly and Company; F. Hoffmann-La Roche Ltd and its
affiliated company Genentech, Inc.; GE Healthcare; Innogenetics, N.V.; IXICO Ltd.; Janssen Alzheimer
Immunotherapy Research $\&$ Development, LLC.; Johnson $\&$ Johnson Pharmaceutical Research $\&$
Development LLC.; Medpace, Inc.; Merck $\&$ Co., Inc.; Meso Scale Diagnostics, LLC.; Novartis
Pharmaceuticals Corporation; Pfizer Inc.; Servier; Synarc Inc.; and Takeda Pharmaceutical Company. The
Canadian Institutes of Health Research is providing funds to support ADNI clinical sites in Canada. Private sector contributions are facilitated by the Foundation for the National Institutes of Health (www.fnih.org).
The grantee organization is the Northern California Institute for Research and Education, and the study is
coordinated by the Alzheimer's Disease Cooperative Study at the University of California, San Diego. ADNI
data are disseminated by the Laboratory for Neuro Imaging at the University of California, Los Angeles. This
research was also supported by NIH grants P30 AG010129 and K01 AG030514.

\bibliographystyle{abbrvnat}
\bibliography{refs}

\newpage

\begin{table}[!p]
\centering
\begin{tabular}{c c c c}
\hline
  & Accept null & Reject null & Total \\ [0.5ex]
    &hypothesis &hypothesis &  \\ [0.5ex]
\hline
Null true & $U$ & $V$ & $m_0$\\
Alternative true & $T$ & $S$ & $m_1$ \\
              & $W$ & $R$ & $m$\\
\hline %inserts single line
\end{tabular}
\caption{\emph{Summary of the possible outcomes for all $m$ hypotheses.}}
\label{tab:mtests}
\end{table}

\begin{table}[h]
\centering
\begin{tabular}{c c c c } \hline\hline
Size	&	Simulation 1	&	Simulation 2	&	Simulation 3	\\
\hline
0.1	&	0.115	&	0.110	&	0.110	\\
0.2	&	0.219	&	0.221	&	0.224	\\
0.3	&	0.309	&	0.324	&	0.326	\\
0.4	&	0.392	&	0.415	&	0.416	\\
0.5	&	0.472	&	0.498	&	0.499	\\
0.6	&	0.556	&	0.576	&	0.578	\\
0.7	&	0.653	&	0.656	&	0.658	\\
0.8	&	0.775	&	0.746	&	0.748	\\
0.9	&	0.940	&	0.878	&	0.877	\\
1.0	&	1.000	&	1.000	&	1.000	\\
\hline
\end{tabular}
\caption{\emph{The empirical size results from the three simulations for the size analysis.}}
\label{tab:sizeanalysis}
\end{table}

\begin{landscape}
\begin{table}[!p]
\centering
\begin{tabular}{c c c c c c c c} \hline \hline
\multicolumn{2}{c}{ } &\multicolumn{2}{c}{Alg. 1: $p$values+$q$value} &\multicolumn{2}{c}{Alg. 2: $p$values+local fdr}  &\multicolumn{2}{c}{Alg. 3: local fdr modeling} \\ \hline
\multicolumn{1}{c}{ } & \multicolumn{1}{c}{$\alpha$ }& \multicolumn{1}{c}{FDR (s.e.) } & \multicolumn{1}{c}{power (s.e.)} & \multicolumn{1}{c}{FDR (s.e.) } & \multicolumn{1}{c}{power (s.e.)}& \multicolumn{1}{c}{FDR (s.e.) } & \multicolumn{1}{c}{power (s.e.)}\\ \hline
\multirow{4}{*}{Simulation 1}
&	0.05	&	0.000	(0.000)	&	0.000	(0.000)	&	0.019	(0.026)	&	0.882	(0.065)	&	0.039	(0.121)	&	0.904	(0.068)	\\
&	0.10	&	0.006	(0.011)	&	0.691	(0.114)	&	0.063	(0.050)	&	0.964	(0.032)	&	0.080	(0.122)	&	0.965	(0.033)	\\
&	0.15	&	0.007	(0.012)	&	0.777	(0.142)	&	0.121	(0.068)	&	0.985	(0.019)	&	0.128	(0.127)	&	0.984	(0.020)	\\
&	0.20	&	0.028	(0.076)	&	0.900	(0.085)	&	0.186	(0.080)	&	0.993	(0.013)	&	0.184	(0.126)	&	0.990	(0.029)	\\

 \hline
\multirow{4}{*}{Simulation 2}
&   0.05	&	0.000	(0.000)	&	0.000	(0.000)	&	0.020	(0.026)	&	0.801	(0.106)	&	0.035	(0.044)	&	0.882	(0.087)	\\
&   0.10	&	0.000	(0.000)	&	0.035	(0.026)	&	0.071	(0.051)	&	0.916	(0.085)	&	0.088	(0.061)	&	0.934	(0.038)	\\
&   0.15	&	0.002	(0.007)	&	0.467	(0.175)	&	0.134	(0.070)	&	0.950	(0.064)	&	0.150	(0.075)	&	0.959	(0.025)	\\
&   0.20	&	0.008	(0.016)	&	0.665	(0.159)	&	0.194	(0.080)	&	0.964	(0.048)	&	0.212	(0.082)	&	0.970	(0.016)	\\
 \hline
\multirow{4}{*}{Simulation 3}
&	0.05	&	0.000	(0.000)	&	0.000	(0.000)	&	0.017	(0.024)	&	0.758	(0.095)	&	0.030	(0.033)	&	0.844	(0.066)	\\
&	0.10	&	0.000	(0.000)	&	0.020	(0.000)	&	0.064	(0.051)	&	0.899	(0.059)	&	0.084	(0.070)	&	0.918	(0.045)	\\
&	0.15	&	0.002	(0.007)	&	0.418	(0.179)	&	0.126	(0.070)	&	0.944	(0.036)	&	0.145	(0.091)	&	0.951	(0.029)	\\
&	0.20	&	0.004	(0.011)	&	0.572	(0.159)	&	0.194	(0.084)	&	0.963	(0.024)	&	0.208	(0.103)	&	0.965	(0.029)	\\
 \hline
\end{tabular}
\caption{\emph{Simulation results of the average FDR estimates (with standard error) and the average power estimates (with standard error) of the three algorithms described in Section \ref{Multiple testing procedure}. Simulation 1 is based on linear correlation; simulation 2 is based on mixed correlations with independent covariance matrix; and simulation 3 is based on mixed correlations with dependent covariance matrix in Section \ref{Simulation Design}.}}
\label{tab:pFDRRresult}
\end{table}
\end{landscape}

\begin{landscape}
\begin{table}[!p]
\centering
\begin{tabular}{c c c c c } \hline \hline
\multicolumn{1}{c}{ $\alpha$ } &\multicolumn{1}{c}{Alg. 1: $p$values+$q$value} &\multicolumn{1}{c}{Alg. 2: $p$values+local fdr}  &\multicolumn{1}{c}{Alg. 3: local fdr modeling} &\multicolumn{1}{c}{Alg. 4: slr+local fdr} \\ \hline
 0.05 	 & 	 5,388 	 & 	 27,965 	 & 	 23,128 	 & 	 0   	 \\
 0.10 	 & 	 8,447 	 & 	 34,659 	 & 	 29,288 	 & 	 18 	 \\
 0.15 	 & 	 11,041 	 & 	 39,875 	 & 	 34,261 	 & 	 38 	 \\
 0.20 	 & 	 13,804 	 & 	 44,604 	 & 	 38,794 	 & 	 95 	 \\
 0.30 	 & 	 19,299 	 & 	 53,716 	 & 	 47,535 	 & 	 275 	 \\
 0.40 	 & 	 25,537 	 & 	 63,449 	 & 	 56,853 	 & 	 612 	 \\
 0.50 	 & 	 448,073 	 & 	 75,030 	 & 	 68,365 	 & 	 1,180 	 \\
 \hline
\end{tabular}
\caption{\emph{Number of significant SNPs found by the different algorithms with different $\alpha$ levels, under the region-wide study. The ADNI study was used for this experiment.}} %The nominal alpha level was set up to 0.05.}}
\label{tab:ADNIresult}
\end{table}
\end{landscape}

\begin{table}[!p]
\centering
\begin{tabular}{c c c c c c} \hline \hline
\multicolumn{5}{c}{Top 1180 SNPs from each algorithm in table \ref{tab:ADNIresult}} \\ \hline
Annotation cluster & alg. 1$^a$  & alg. 2$^b$ & alg. 3$^c$ & alg. 4$^d$ \\ \hline
%\multicolumn{1}{c}{Functional Annotation Clustering}& \multicolumn{1}{c}{$w_{1,0.05}$} & \multicolumn{1}{c}{$r_{1,0.05}$} & \multicolumn{1}{c}{$w_{1,0.4}$} & \multicolumn{1}{c}{$r_{1,0.4}$}& \multicolumn{1}{c}{$w_{2,0.05}$} & \multicolumn{1}{c}{$r_{2,0.05}$} & \multicolumn{1}{c}{$w_{3,0.05}$} & \multicolumn{1}{c}{$r_{3,0.05}$}\\ \hline
%annotation cluster	 & 	1180/5388	 & 	1180/27965	 & 	1180/23128	 & 	1180	\\
1	 & 	 3.167 	 & 	 3.167 	 & 	 3.167 	 & 	 1.479 	 \\
2	 & 	 1.680 	 & 	 1.680 	 & 	 1.680 	 & 	 1.332 	 \\
3	 & 	 1.198 	 & 	 1.198 	 & 	 1.198 	 & 	 1.040 	 \\
4	 & 	 1.157 	 & 	 1.157 	 & 	 1.157 	 & 	 0.775 	 \\
5	 & 	 1.014 	 & 	 1.014 	 & 	 1.014 	 & 	 0.508 	 \\
6	 & 	 0.947 	 & 	 0.947 	 & 	 0.947 	 & 	 0.261 	 \\
7	 & 	 0.572 	 & 	 0.572 	 & 	 0.572 	 & 	 0.175 	 \\
8	 & 	 0.523 	 & 	 0.523 	 & 	 0.523 	 & 	 0.169 	 \\
total	 & 	 10.257 	 & 	 10.257 	 & 	 10.257 	 & 	 5.739 	 \\
 \hline
\end{tabular}
\caption{\emph{Enrichment scores for each cluster from DAVID database. $^a$: top 1180 SNPs were selected from 5388 SNPs in algorithm 1 at level 0.05; $^b$: top 1180 SNPs were collected from 27965 SNPs in algorithm 2 at level 0.05; and $^c$: top 1180 SNPs were collected from 23128 SNPs in algorithm 3 at level 0.05; $^d$: 1180 SNPs found at level 0.5 in algorithm 4.}}
\label{tab:David}
\end{table}

\end{document}